\documentclass[12pt]{JHEP3}
\usepackage{amsfonts}
\usepackage{bbold}

\usepackage{amsmath}

\title{Gravitational F-terms of SO/Sp gauge theories and anomalies}
\preprint{\hepth{0306299}\\SISSA 61/2003/EP}
\author{L. F. Alday$^{1,2}$ ,M. Cirafici$^{1}$
\\
$^1$International School for Advanced Studies, Trieste, Italy and I.N.F.N. sez. di Trieste.\\
$^2$The Abdus Salam ICTP, Trieste, Italy.\\
Email: \, \email{alday@sissa.it},\, \email{cirafici@sissa.it}}

\abstract{We study the first non trivial gravitational corrections to the F-terms of ${\cal N}=1$ SYM SO/Sp gauge theories, with matter in some representations, by using the generalized Konishi anomaly method. We derive equations at genus one for the operators in the chiral ring and compare them with the loop equations of the corresponding matrix models, finding agreement. We find that for adjoint representation the genus 0 contributions to such corrections can be adsorbed by a field redefinition; remarkably, this is not the case for matter in the (anti-)symmetric representation of ($Sp$) $SO$.}

\begin{document}

\section{Introduction}

The dynamics of ${\cal N}=1$ gauge theories has been better understood since the Dijkgraaf and Vafa (DV) conjecture \cite{Dijkgraaf:2002dh}, relating the effective superpotential of a wide class of ${\cal N}=1$ gauge theories with the free energy of related matrix models; since then, many checks of the conjecture have been done, for instance it has been applied to a variety of gauge theories, by using perturbative arguments, and by using anomalies \cite{literature} \cite{Cachazo:2002ry} \cite{Alday:2003gb} \cite{Kraus:2003jv} \cite{Feng:2002gb} \cite{Janik:2002nz}. In particular, DV conjectured that for gauge theories in flat space-time the effective superpotential of such gauge theory can be computed by summing planar diagrams in the related matrix model, whose action is given (up to a factor) by the superpotential of the gauge theory. If one, instead, considers gauge theories in non flat backgrounds, for instance turning on the ${\cal N}=1$ Weyl multiplet or the graviphoton, then the non-planar diagrams of the matrix model become relevant. The precise relation between these gravitational F-terms and the non planar diagrams of the matrix model has been studied in \cite{Dijkgraaf:2002yn} \cite{Klemm:2002pa} \cite{Ooguri:2003tt} \cite{Ooguri:2003qp}. 

The original conjecture made use of String theory, a direct field theory derivation was given in \cite{Cachazo:2002ry}, by using Ward identities from the generalized Konishi anomaly. The issue of gravitational F-terms was studied in \cite{David:2003ke} \cite{Alday:2003ms} by applying this same method.

In \cite{David:2003ke} it was considered the first non trivial gravitational correction to the effective superpotential in a $U(N)$ gauge theory with a single adjoint chiral multiplet, by using the generalized Konishi anomaly and the gravitationally deformed chiral ring. The purpose of this paper is to extend such analysis to $SO/Sp$ gauge theories with matter in both adjoint and symmetric representations (or adjoint and antisymmetric for $Sp$).

As for the $U(N)$ case, the three key points are the gravitational deformation of the chiral ring, the gravitational contribution to the Konishi anomaly and the non factorization of product of operators in the chiral ring.

With these ingredients we derive genus one equations for the relevant operators and compare them with the matrix model loop equations. We show that gravitational corrections get also a genus zero contribution. This contribution can be adsorbed by a field redefinition when matter is in the adjoint representation, but, interestingly, this is not the case when matter is in the (anti-)symmetric representation for ($Sp$) $SO$. This is related to the fact that these theories cannot be obtained by softly breaking ${\cal N}=2$ to ${\cal N}=1$ by turning on fluxes, then the stringy  arguments of \cite{Ooguri:2003tt} \cite{Ooguri:2003qp} does not necessarily apply. 

The present paper is organized as follows: in section 2 we list some preliminaries results that will be used in the rest of the paper, in section 3 we find the genus one gravitational correction corresponding to a $SO(N)$ gauge theory with matter in the adjoint representation and generalize such results for the symmetric representation and for the $Sp(N)$ group, in section 4 we compare our results with the matrix model analysis and finally we end with some discussion and open problems.

\section{Preliminaries}

In this section we set some notation and results that will be useful in the calculation of the genus one contributions to the effective superpotential. We consider a ${\cal N}=1$ gauge theory coupled to  ${\cal N}=1$ supergravity with a matter chiral field $\Phi$.

The chiral ring is spanned by all the chiral operators (annihilated by the covariant derivative $\overline{D}_{\dot{\alpha}}$) modulo $\overline{D}_{\dot{\alpha}}$ exact terms, where the exact terms should be gauge invariant and local operators. In the presence of the ${\cal N}=1$ Weyl multiplet $G_{\alpha\beta\gamma}$  , the chiral ring relations are \cite{Ooguri:2003tt} 
\begin{equation}
\label{chiral}
\left[ W^{\alpha} , \Phi \right]=0, \hspace{0.3in}
\{W_\alpha , W_\beta\} =2 G_{\alpha\beta\gamma}
 W^{\gamma} 
\end{equation} 
The full set of operators in the chiral ring can be encoded in the following operators
\begin{eqnarray}
\label{defoper}
{\cal R}(z)_{ij} = -\frac{1}{32\pi^2} \left( \frac{W^2}{z-\Phi}
\right)_{ij},   &\;& R(z) = {\rm Tr } {\cal R}(z), \nonumber \\ 
\rho_{\alpha}(z)_{ij}=\frac{1}{4 \pi} \left( \frac{W_{\alpha}}{z-\Phi} \right)_{ij}, &\;& w_{\alpha}(z)={\rm Tr} \rho_{\alpha}(z), \\
\nonumber
{\cal T}(z)_{ij} = \left( \frac{1}{z-\Phi} \right)_{ij}, &\;&
T(z) = {\rm Tr} {\cal T} (z). 
\end{eqnarray}
From (\ref{chiral}) one can derive the following identities \footnote{From now on all the identities are intended to hold in the chiral ring, {\it i.e.} 
 $\hbox{mod }\bar{D}$.  }
\begin{eqnarray}
[W_\alpha, W^2] = 0,   &\;& \hspace{0.3in}
\{ W_\alpha, W^2 \} = -\frac{2}{3} G^2 W_\alpha, \\ \nonumber
W_\alpha W^2 =  - \frac{1}{3} G^2 W_\alpha,  \nonumber &\;& \hspace{0.3in}
W^4 = - {1 \over 3} G^2 W^2
\end{eqnarray}
These identities imply that gauge invariant combinations of
certain chiral operators vanish in the chiral ring, for instance

\begin{equation}
\label{gw}
G_{\alpha\beta\gamma} \rm{Tr}( W^\gamma\Phi \Phi \ldots) = 0. \;\;\;\;
\end{equation}
which can also be written as:

\begin{equation}
G_{\alpha \beta \gamma} w^{\gamma}(z) = 0
\end{equation}
Multiplying (\ref{gw}) by $G^{\alpha\beta}_{\delta}$ we obtain
\begin{equation}
G^2 \rm{Tr}( W_\alpha \Phi \Phi \ldots ) = 0.
\end{equation}
Another important identity in the chiral ring is 
\begin{equation}
G^4 = (G^2)^2 =0
\end{equation}
In particular, the corrections due to the ${\cal N} =1$ Weyl multiplet, $G_{\alpha\beta\gamma}$, truncate at order $G^2$. 

Note that in this analysis we are not considering the graviphoton field strength $F_{\alpha\beta}$, such field is necessary if one wants to capture arbitrary genus expansion, since arbitrary powers of $F_{\alpha\beta}$ are non vanishing.

\section{Genus one solution}

In this section we will derive the genus one contributions to the operators (\ref{defoper}). The equations for such operators can be derived by the method of the generalized Konishi anomaly, that is one derives Ward identities in the chiral ring by considering the anomalous variation $\delta \Phi_{ij} = f_{ij}$ which gives rise to the following anomaly
\begin{equation}
\frac{\delta f_{ji}}{\delta \Phi_{k \ell}} A_{ij,k \ell}
\end{equation}

In the following we study the gravitational corrections to the effective superpotential, taking into account the ${\cal N}=1$ Weyl multiplet. We do it for a $SO(N)$ gauge theory with matter in the adjoint representation in detail and then state the results for other groups and representations.

\subsection{$SO(N)$ with matter in the adjoint representation}

For the case of $SO(N)$\footnote{For groups others than $U(N)$, it was explained in \cite{Alday:2003gb} how to derive the anomaly equation, in absence of gravity. See also \cite{Kraus:2003jv} for a similar analysis.}, for matter in the adjoint representation, and in presence of the gravitational background under consideration we have \cite{Magnoli:1989qh} \cite{Konishi:1988mb}
\begin{equation}
A_{ij,k \ell} = (W^2)_{kj} \delta_{i\ell} + \delta_{kj} (W^2)_{i\ell} - 2
W^{\alpha}_{kj} W_{\alpha i\ell} + {1 \over 3} G^2 \delta_{kj}\delta_{i\ell} 
- \left( k \longleftrightarrow \ell \right).
\end{equation}
Now, by taking  variations of the form $\delta \Phi_{ij}= {\cal R}(z)_{ij}$,  
$\delta \Phi_{ij}= \eta^\alpha \rho_{\alpha}(z)_{ij}$ and  
$\delta \Phi_{ij}= {\cal T}(z)_{ij}$, we obtain the following equations
\begin{eqnarray}
\label{rteq}
& &{1 \over 2}\langle R(z) R(z) \rangle - {1 \over 2} G^2 { \langle R(z) \rangle \over z} - \langle {\rm Tr}( V'(\Phi){\cal R}(z)) \rangle   = 0,\nonumber \\
& & \langle R(z) w_{\alpha}(z) \rangle - \langle {\rm Tr}( V'(\Phi) \rho_{\alpha}
(z)) \rangle = 0 \\
& &\langle R(z) T(z) \rangle  -\langle {\rm Tr}(V'(\Phi){\cal T}(z)) \rangle  
\nonumber
   - \frac{1}{6} G^2 \langle T(z) T(z) \rangle \\ \nonumber
& & \,\,\,\,\,\,\,\,\,\,\,\,\,\,  - 2 { \langle R(z) \rangle \over z} 
+  {1 \over 6} G^2 { \langle T(z) \rangle \over z} + {1 \over 2} \langle w_{\alpha}(z) w^{\alpha}(z) \rangle = 0.
\end{eqnarray}
where $V(\Phi)$ is the tree level superpotential. As for the $U(N)$ case, two point functions do not factorize because of the non trivial gravitational background, but in general contain a non vanishing connected piece. Then we need a set of equations for the connected two point functions; such equations are obtained by considering variations of the form
\begin{equation}
\delta \Phi_{ij} = {\cal A}_{ij}(z) B(w)
\end{equation}
where ${\cal A}$ can be  ${\cal R}$, $\eta^\alpha \rho_{\alpha}$ or ${\cal T}$ and $B$ can be $R$, $\eta^\alpha w_{\alpha}$ or $T$. The reason for considering two point functions in different points $w$ and $z$ is that we can impose conditions on them that their integrals around various branch cuts in $w$ and $z$ planes vanish separately. 
The Jacobian of this transformation has two pieces
\begin{eqnarray}
\frac{\delta (\delta \Phi_{ji})}{\delta \Phi_{kl}} &=& \frac{\delta
{\cal A}_{ji}(z)}
{\delta \Phi_{kl}} B(w) + \sum_m  {\cal A}_{ji}(z) \left( {\cal B}_{mk}(w) {\cal T}_{lm}(w) - {\cal B}_{ml}(w) {\cal T}_{km}(w) \right)\\
&=& \frac{\delta
{\cal A}_{ji}(z)}
{\delta \Phi_{kl}} B(w) + \sum_m  {\cal A}_{ji}(z) \left( {\cal B}_{mk}(w) {\cal T}_{lm}(w) \mp {\cal B}_{mk}(-w) {\cal T}_{lm}(-w) \right)\nonumber
\end{eqnarray}
where the plus sign holds when ${\cal B}_{mk}(w)= \eta^\alpha \rho_\alpha(w)_{mk} $ and the minus otherwise, this can be seen by using the symmetry properties of the field $\Phi$, which give
\begin{equation}
T(w)^{t}=-T(-w), \hspace{0.6in} w_{\alpha}(w)^t=w_{\alpha}(-w), \hspace{0.6in} R(w)^t=-R(-w)
\end{equation}
With $T(z)^t$ we mean ${\rm Tr} {\cal T}^t (z)$, etc. Considering these variations together with the classical variation of the superpotential, one obtains a set of equations for the two point connected functions that can be summarized in the following matrix equation
\begin{eqnarray}
&~&
\left[   \begin{array}{ccc}
M(z)  &
\langle T(z)\rangle - {2 \over z}
 & 0 \\
0 & M(z) &0 \\
0 & 0 & M(z)                       \end{array} \right]
        \left[
\begin{array}{ccc}
\langle T(z)T(w)\rangle_c &\langle T(z)R(w)\rangle_c &\langle T(z)
w_{\beta}(w)\rangle_c \\
\langle R(z)T(w)\rangle_c &\langle R(z)R(w)\rangle_c &\langle R(z)
w_{\beta}(w)\rangle_c \\
\langle w_{\alpha}(z)T(w)\rangle_c &\langle w_{\alpha}(z)R(w)\rangle_c 
&\langle w_{\alpha}(z)
w_{\beta}(w)\rangle_c \end{array}\right]
= \nonumber\\ &~& =
\frac{G^2}{6} \partial_w
\left[
\begin{array}{ccc}
\tilde{T}(z,w) & \tilde{R}(z,w) & 0\\
\tilde{R}(z,w) & 0 & 0 \\
0 & 0 & 5\epsilon_{\alpha \beta} \tilde{R}(z,w) \end{array}\right]
\label{matrix}
\end{eqnarray}
where we have introduced the operator $M(z) = \langle R(z) \rangle - I(z)$ and $\tilde{B}(z,w)=B(z,w)+B(z,-w)$ where 
\begin{equation}
B(z,w)= {\langle B(z) \rangle - \langle B(w) \rangle \over z - w }
\end{equation}
Here the integral operator $I(z)$  is given by
\begin{equation}
I(z) A(z) = \frac{1}{2\pi i} \oint _{C_z} dy \frac{ V'(y) A(y)}{y-z} ,
\end{equation}
with the contour $C_z$ encircling $z$ and $\infty$. Note that if $A$
is equal to ${\cal R} $, $\eta^\alpha \rho_{\alpha}$ or ${\cal T}$, the integral operator reduces to
\begin{equation}
I(z) A(z) = V'(\Phi) A(z)
\end{equation}
The  equation in (\ref{matrix}) is of the form
\begin{equation}
{\cal M}(z) N(z,w) =\partial_w  K(z,w)
\end{equation}
with ${\cal M}(z)$  the matrix operator appearing on the left hand
side of eq. (\ref{matrix}), and $N(z,w)$ and $K(z,w)$ satisfying $N(z,w)=
N^t(w,z)$ and $K(z,w) = K^t(w,z)$. The non-trivial consistency
condition (integrability condition) is then given by
\begin{equation}
(\partial_w K(z,w)) {\cal M}^t(w) = {\cal M}(z) \partial_z K(z,w).
\label{int}
\end{equation}
Using the methods of \cite{David:2003ke}, it can be shown that the
above integrability condition is satisfied, the existence of solutions for the connected two point functions, equation 
(\ref{matrix}), is then guaranteed. However these solutions suffer from
ambiguities, in the form of a finite set of parameters. 
These ambiguities will be fixed by the physical requirement that the contour 
integrals around 
the branch cuts of the connected 
two point-functions, both in the $z$ and $w$ planes, vanish separately.
The reason for this is that the following operator equations hold:
\begin{equation}
\label{opequ}
\frac{1}{2\pi i}\oint_{C_i}dz R(z) = S_i, ~~~~ \frac{1}{2\pi i}
\oint_{C_i} dw T(w) = N_i, ~~~~\frac{1}{2\pi i}\oint_{C_i}dz w_{\alpha}(z) =
w_{\alpha~i}.
\end{equation}
where $S_i$ is the chiral superfield whose lowest component is the
gaugino bilinear in the $i$-th gauge group factor in the broken phase
$SO(N) \rightarrow  SO(N_0) \times \prod_{i=1}^n U(N_i)$ and 
$w_{\alpha~i}$ is the $U(1)$ chiral gauge superfield of the $U(N_i)$
subgroup. Since these fields are background fields, in the connected
correlation functions the contour integrals around the branch cuts
must vanish: 

\begin{equation}
\oint_{C_i}dz \langle A(z) B(w) \rangle_c = \oint_{C_j}dw \langle A(z) B(w) \rangle_c=0
\end{equation}

This requirement makes the solution of the equations of
(\ref{matrix}) unique. We can express such solutions in term of one function $H(z,w)$, whose explicit form will be not needed for the analysis done here. The two point functions we will need are
\begin{eqnarray}
\langle R(z) R(w) \rangle _c &=& 0 \nonumber\\\langle R(z) T(w) \rangle _c &=& {1 \over 6} G^2 H(z,w)  \\\hspace{0.3in} \langle w(z)_{\alpha} w_{\beta}(w) \rangle _c&=&{5 \over 6} G^2 H(z,w) \epsilon_{\alpha \beta}\nonumber
\end{eqnarray}
Now we could plug these solutions into (\ref{rteq}) and solve for the one point functions. However note that we can instead perform the following rescaling \footnote{Note that such rescaling changes the boundary conditions (\ref{opequ}) in a non trivial way .}
\begin{equation}
\label{redef}
R(z) \rightarrow R(z) + \frac{G^2}{6} T(z) + \frac{G^2}{6} \frac{1}{z}
\end{equation}
With this, equations (\ref{rteq}) become
\begin{eqnarray}
\label{rteqresc}
& &{1 \over 2}\langle R(z) R(z) \rangle -\langle {\rm Tr}( V'(\Phi){\cal R}(z)) \rangle   = 0,\\
& &\langle R(z) T(z) \rangle  -\langle {\rm Tr}(V'(\Phi){\cal T}(z)) \rangle  
\nonumber
 - 2 { \langle R(z) \rangle \over z} -\frac{1}{3}\frac{G^2}{z^2}+{1 \over 2}\langle w_{\alpha}(z) w^{\alpha}(z) \rangle =0.
\end{eqnarray}
Where the chiral ring relations and equations of motion have been used. We have not included the equation for $w_{\alpha}(z)$ since it can be set consistently to zero. In order to find the corrections to the zero order one point functions, we expand as follows
\begin{equation}
\langle R(z) \rangle = R^{(0)}(z) +  R^{(1)}(z), \;\;
\langle T(z) \rangle = T^{(0)}(z) +  T^{(1)}(z).
\label{exp}
\end{equation}
With the superscript 0 with denote zero order in $G^2$ and with the superscript 1, order one in $G^2$. Note that, since $G^4$ terms are trivial in the chiral ring, this expansion is exact. 
Substituting the above expansions in (\ref{rteqresc}) we obtain for order 0
\begin{eqnarray}
{1 \over 2} R^{(0)}(z) R^{(0)}(z) - I(z) R^{(0)}(z) = 0 \nonumber\\
 R^{(0)}(z) T^{(0)}(z) - I(z) T^{(0)}(z) - {2 \over z}  R^{(0)}(z) = 0
\end{eqnarray}
Note that these are the same equations already obtained in \cite{Alday:2003gb}.
For order one, we find
\begin{eqnarray}
M(z) R^{(1)}(z) &=& 0 \nonumber \\
M(z) T^{(1)}(z) &=& {1 \over 3} G^2 {1 \over z^2} - G^2 H(z,z) 
\end{eqnarray}
Note that, as expected, the genus zero contribution ({\it i.e.} going with $N^2$) has been adsorbed with the redefinition (\ref{redef}).

The results of this subsection are for a $SO(N)$ gauge theory with matter in the adjoint representation, in the following we show the extension of such results for the symmetric representation, as for the $Sp(N)$ gauge group with matter in both adjoint and antisymmetric representations.

\subsection{$SO(N)$ with matter in the symmetric representation}

For the case of a $SO(N)$ gauge group with matter in the symmetric representation we have
\begin{equation}
A_{ij,k \ell} = (W^2)_{kj} \delta_{i\ell} + \delta_{kj} (W^2)_{i\ell} - 2
W^{\alpha}_{kj} W_{\alpha i\ell} + {1 \over 3} G^2 \delta_{kj}\delta_{i\ell} 
+ \left( k \longleftrightarrow \ell \right).
\end{equation}
By using the generalized Konishi anomaly, one can easily obtain the following equations
\begin{eqnarray}
\label{symeq}
& &{1 \over 2}\langle R(z) R(z) \rangle - {1 \over 2} G^2 {\langle \frac{\partial R(z)}{\partial z} \rangle}  - \langle {\rm Tr}( V'(\Phi){\cal R}(z)) \rangle   = 0, \nonumber \\
& & \langle R(z) w_{\alpha}(z) \rangle - \langle {\rm Tr}( V'(\Phi) \rho_{\alpha}
(z)) \rangle = 0 \\
& &\langle R(z) T(z) \rangle  -\langle {\rm Tr}(V'(\Phi){\cal T}(z)) \rangle  
\nonumber
   - \frac{1}{6} G^2 \langle T(z) T(z) \rangle \\ \nonumber
& & \hspace{1in} - 2 { \langle \frac{\partial R(z)}{\partial z} \rangle} 
+  {1 \over 6} G^2 { \langle  \frac{\partial T(z)}{\partial z} \rangle} + {1 \over 2} \langle w_{\alpha}(z) w^{\alpha}(z) \rangle= 0.
\end{eqnarray}
Note that, unlike as for the previous case, the genus 0 contributions cannot be re adsorbed by a field redefinition. The existence of such redefinition is expected for theories that can be obtained by softly breaking ${\cal N}=2$ to ${\cal N}=1$, however note that this is not the case for a $SO(N)$ gauge theory with matter in the symmetric representation.

Again the two point functions will contain a non vanishing connected piece, and we need equations for such pieces. One can easily obtain
\begin{eqnarray}
&~&
\left[   \begin{array}{ccc}
M(z)  &
\langle T(z)\rangle -  2 \frac{\partial}{\partial z}
 & 0 \\
0 & M(z) &0 \\
0 & 0 & M(z)                       \end{array} \right]
        \left[
\begin{array}{ccc}
\langle T(z)T(w)\rangle_c &\langle T(z)R(w)\rangle_c &\langle T(z)
w_{\beta}(w)\rangle_c \\
\langle R(z)T(w)\rangle_c &\langle R(z)R(w)\rangle_c &\langle R(z)
w_{\beta}(w)\rangle_c \\
\langle w_{\alpha}(z)T(w)\rangle_c &\langle w_{\alpha}(z)R(w)\rangle_c 
&\langle w_{\alpha}(z)
w_{\beta}(w)\rangle_c \end{array}\right]
= \nonumber\\ &~& =
\frac{1}{3}G^2 \partial_w
\left[
\begin{array}{ccc}
T(z,w) & R(z,w) & 0\\
R(z,w) & 0 & 0 \\
0 & 0 & 5 R(z,w) \epsilon_{\alpha \beta} \end{array}\right]
\end{eqnarray}
As before, the solution to such equations is unique, and can be expressed in terms of one function $J(z,w)$, the two point functions of interest are
\begin{eqnarray}
\langle R(z) R(w) \rangle _c &=& 0 \nonumber \\
\langle R(z) T(w) \rangle _c &=& {1 \over 3} G^2 J(z,w) \\ 
\langle w_{\alpha}(z) w_{\beta}(w) \rangle _c &=& {5 \over 3} G^2 J(z,w) \epsilon_{\alpha \beta} \nonumber
\end{eqnarray}

Inserting this into (\ref{symeq}) and expanding as before we obtain \footnote{One can see that the zero order equations are the ones obtained in \cite{Alday:2003gb} in absence of gravity}
\begin{eqnarray}
M(z)R^{(1)}(z)&=&{1 \over 2} G^2 \frac{\partial R^{(0)}}{\partial z} \nonumber\\
M(z)T^{(1)}(z)&=&T^{(0)}(z)R^{(1)}(z)+\frac{G^2}{6}(T^{(0)})^2+\\ 
&2& \frac{\partial R^{(1)}(z)}{\partial z} -\frac{G^2}{6}\frac{\partial T^{(0)}(z)}{\partial z}-{1 \over 3} G^2 J(z,w) \nonumber
\end{eqnarray}

\subsection{$Sp(N)$ gauge theory}

Now we will focus on an $Sp(N)$ gauge theory with matter in the adjoint 
(symmetric) and in the antisymmetric  
representations. With symmetric (antisymmetric) we 
mean that $\Phi$ has to be considered as a matrix $MJ$ where $M$ is a 
symmetric (antisymmetric) matrix and $J$ is the invariant antisymmetric 
tensor of $Sp(N)$. We take the generators of $Sp(N)$ to be $ 
\left(e_{lk}+e_{kl} \right)$ with $(e_{lk})_{ij}=\delta_{il} \delta_{jk}$. The 
analysis for the $Sp(N)$ case is almost identical to the one for the $SO(N)$ 
case, the only change being the sign in the generators (and of course the 
different properties of the matrices representing the field $\Phi$, since the 
antisymmetric invariant $J$ will enter in the intermediate steps). Because of
 this the only difference with the case of $SO(N)$ will be a change of sign in front of the terms ${1 \over z}$ for matter in the adjoint representation and of the terms ${ \partial R(z) \over {\partial z}}$ and ${\partial T(z) \over {\partial z}}$ for antisymmetric matter. The reason for this was explained in 
\cite{Alday:2003gb}.
The results obtained are
\begin{eqnarray}
M(z) R^{(1)}(z) &=& 0 \nonumber\\
M(z) T^{(1)}(z) &=& {1 \over 3} G^2 {1 \over z^2} - G^2 H^{Sp}(z,z) 
\end{eqnarray}
for the adjoint representation (note that the genus zero contributions have been absorbed) and
\begin{eqnarray}
M(z)R^{(1)}(z)&=&-{1 \over 2} G^2 \frac{\partial R^{(0)}}{\partial z} \nonumber \\
M(z)T^{(1)}(z)&=&T^{(0)}(z)R^{(1)}(z)+\frac{G^2}{6}(T^{(0)})^2\\
&-&2 \frac{\partial R^{(1)}(z)}{\partial z} +\frac{G^2}{6}\frac{\partial T^{(0)}(z)}{\partial z}-{1 \over 3} G^2 J^{Sp}(z,z) \nonumber
\end{eqnarray}
for matter in the antisymmetric representation.

\section{Comparison with matrix model}

In this section we will show how the kind of relations obtained previously can be compared with the matrix model results. In particular, we will focus on the case of $SO(N)$ with adjoint and symmetric matter (an analogous treatment can also be done for $Sp(N)$). In the case of planar diagrams, the matrix model for $SO(N)$ was studied in \cite{Feng:2002gb} \cite{Janik:2002nz} \cite{Ashok:2002bi} 
\cite{Fuji:2002wd}.

\subsection{$SO(N)$ with adjoint matter}

Let us consider the following one--matrix model with action

\begin{equation}
S={M \over g_s} \sum_k {g_k \over M} {\rm tr} \Phi^k
\end{equation}
where $\Phi$ a $M \times M$ matrix in the adjoint representation of $SO(M)$. The resolvent 
\begin{equation}
\Omega(z)= {g_s \over M} {\rm tr} {1 \over z - \Phi}
\end{equation}
satisfies the following (loop) equation
\begin{equation}
{1 \over 2} \langle \Omega(z)\rangle^2 -I(z) \langle \Omega(z) \rangle 
 - {1 \over M} {1 \over 2z} \langle \Omega(z) \rangle + \frac{1}{2} \langle \Omega(z) \Omega(z) \rangle_c = 0
\end{equation}
remember that the connected part of the correlator $\langle \Omega(z) \Omega(z) \rangle_c$ goes like ${1 \over M^2}$ and is usually neglected when looking at the planar limit (large $M$ limit); however for the analysis done here it is needed. In order to show the $M$ dependence explicitly, we will write it as 
${1 \over 2} \langle \Omega(z) \Omega(z) \rangle_c = {1 \over M^2} \Omega(z,z)$. Note that this loop equation is analogous to the equation fulfilled by the generator $R(z)$ in the corresponding gauge theory, equation (\ref{rteq}), provided we identify $G^2$ with ${1 \over M}$.

Next we expand the resolvent in powers of ${1 \over M}$ as
\begin{equation}
\label{expomega}
\langle \Omega(z) \rangle = \Omega^{(0)}(z)+ {1 \over M} \Omega^{(1)}(z) + 
 {1 \over M^2}\Omega^{(2)}(z) + \cdots
\end{equation}
Plugging (\ref{expomega}) into the loop equation, one finds
\begin{eqnarray}
\label{mmequations}
\left( {1 \over 2} \Omega^{(0)}(z) - I(z) \right) \Omega^{(0)}(z) = 0 \nonumber \\
\label{secondeq}
\left( \Omega^{(0)}(z) - I(z) \right) \Omega^{(1)}(z) - {1 \over 2z} \Omega^{(0)}(z) = 0 \\
\left( \Omega^{(0)}(z) - I(z) \right) \Omega^{(2)}(z) - {1 \over 2z} \Omega^{(1)}(z) + {1 \over 2} ( \Omega^{(1)}(z) )^2 + \Omega(z,z) = 0 \nonumber
\end{eqnarray}

In order to compare these equations with their gauge theory counterpart let us perform the following rescaling 
\begin{equation}
\label{pr}
R(z)\rightarrow R(z) + \frac{G^2}{6}T(z)
\end{equation} 
Under this, equations (\ref{rteq}) become
\begin{eqnarray}
& &{1 \over 2}\langle R(z) R(z) \rangle - {1 \over 6} G^2 { \langle R(z) \rangle \over z} - \langle {\rm Tr}( V'(\Phi){\cal R}(z)) \rangle   = 0,\nonumber \\
& &\langle R(z) T(z) \rangle  -\langle {\rm Tr}(V'(\Phi){\cal T}(z)) \rangle  
   \\ \nonumber
& & \,\,\,\,\,\,\,\,\,\,\,\,\,\,  - 2 { \langle R(z) \rangle \over z} 
-  {1 \over 6} G^2 { \langle T(z) \rangle \over z} + {1 \over 2} \langle w_{\alpha}(z) w^{\alpha}(z) \rangle = 0.
\end{eqnarray}
We can see that $\Omega^{(0)}$ and $R^{(0)}$ follow the same equations, and the same happens for $\Omega^{(1)}$ and $R^{(1)}$. Let us now define the following operator:
\begin{equation}
D=\sum_{i}N_{i}\frac{\partial}{\partial S_i}
\end{equation}
Introducing the following notation
\begin{equation}
T^{(0)}(z)=D R^{(0)}(z) + \hat{T}^{(0)}(z),\hspace{0.3in}T^{(1)}(z)=D R^{(1)}(z) + \hat{T}^{(1)}(z) 
\end{equation}
We see that $\hat{T}^{(0)}(z)$ and $\hat{T}^{(1)}(z)$ satisfy the following equations
\begin{eqnarray}
M(z)\hat{T}^{(0)}(z)-\frac{2}{z}R^{(0)}(z)=0\\
M(z)\hat{T}^{(1)}(z)-\frac{2}{z}R^{(1)}(z)+\hat{T}^{(0)}(z)R^{(1)}(z)-\frac{G^2}{6}\frac{1}{z}\hat{T}^{(0)}(z)+connected=0
\end{eqnarray}
With connected we mean connected two point functions. With this we see that $\hat{T}^{(0)}(z)$ and $\hat{T}^{(1)}(z)$ satisfy (up to some factors) the same equations that $\Omega^{(1)}$ and $\Omega^{(2)}$. Note that, for a precise identification, not only the equations but also the boundary conditions (\ref{opequ}) should be the same, that is the reason for using $\hat{T}^{(0)}(z)$ and $\hat{T}^{(1)}(z)$ instead of ${T}^{(0)}(z)$ and $T^{(1)}(z)$.
 
The results obtained here are consistent with the diagrammatic analysis. In calculating the effective superpotential a given diagram of $L$ quantum loops, contributes with a power of $W_{\alpha}$ equal to $2h +4g+2c-2$ with $h$ the number of index loops, $g$ the genus of the diagram and $c$ the number of crosscaps.\footnote{Remember that $L=h+2g+c-1$}. On the other hand, the maximun number of  $W_{\alpha}$ depends on the chiral ring.  In absence of gravitation
\begin{equation}
2h +4g+2c-2 \le 2h  \Rightarrow 4g+2c-2 \le 0
\end{equation}

Implying that only diagrams with $g=c=0$ and $g=0$, $c=1$ will contribute to the effective superpotential, as found for instance in \cite{Alday:2003gb}. In presence of the gravitational background under consideration \footnote{For non trivial gravitational backgrounds we can put more than 2 $W_{\alpha}$ in one index loop, see (\ref{chiral})}

\begin{equation}
2h +4g+2c-2 \le 2h+2  \Rightarrow 4g+2c \le 4 
\end{equation} 

Now diagrams with $g=1$, $c=0$ and $g=0$, $c=2$ will contribute. In order to study which diagrams will contribute to $R(z)$ one should take into account that it contains already two $W_{\alpha}$. 

\subsection{$SO(N)$ with symmetric matter}

In this case, the analysis is completely analogous to the previous one; 
now in the matrix model $\Phi$ is in the symmetric representation of $SO(M)$. The loop equation for the resolvent turns out to be
\begin{equation} 
{1 \over 2} \langle \Omega(z)\rangle^2 -I(z) \langle \Omega(z) \rangle 
 - {1 \over M} \langle {\partial \over \partial z} \Omega(z) \rangle + \frac{1}{2}\langle \Omega(z) \Omega(z) \rangle_c = 0
\end{equation}

Again, note that this loop equation is analogous to the equation fulfilled by the generator $R(z)$ in the corresponding gauge theory, equation (\ref{symeq}), provided we identify $G^2$ with ${1 \over M}$.

Expanding as in (\ref{expomega}) and plugging the expansion in the loop equation, one can obtain
\begin{eqnarray}
\label{mmequationssym}
\left( {1 \over 2} \Omega^{(0)}(z) - I(z) \right) \Omega^{(0)}(z) = 0 \\
\left( \Omega^{(0)}(z) - I(z) \right) \Omega^{(1)}(z) - {\partial \over \partial z} \Omega^{(0)}(z) = 0 \\
\left( \Omega^{(0)}(z) - I(z) \right) \Omega^{(2)}(z) - {\partial \over \partial z} \Omega^{(1)}(z) + {1 \over 2} ( \Omega^{(1)}(z) )^2 + \Omega(z,z) = 0
\end{eqnarray}
The same rescaling (\ref{pr}) can be done here and again the same analysis following equations (\ref{mmequations}) holds.

\section{Conclusions}

In this paper we used the method of the generalized Konishi anomaly in order to compute the first non-trivial gravitational corrections to the F-terms of $SO/Sp$ gauge theories coupled to gravity. 

We analyzed the cases of matter in the adjoint and symmetric representation for $SO$ (anti-symmetric for $Sp$). We found that for the case of adjoint matter the genus zero contributions to gravitational F-terms can be absorbed by a field redefinition, as expected. Interestingly, that is not the case for matter in the symmetric representation, note that a ${\cal N}=1$ $SO(N)$ gauge theory with matter in the symmetric representation cannot be obtained by softly breaking ${\cal N}=2$ to ${\cal N}=1$.

We studied the non-planar diagrams in the dual matrix models, showing that the resolvent of such matrix models encodes the information of the generators in the gauge theory side.

As an open problem will be interesting to include the graviphoton contribution, which will allow to see all genus contributions on the matrix theory side. We stress that the graviphoton is projected out by the orientifold fixed plane needed to engineer the $SO(N)$ gauge theory, and will be absent in the case of unbroken gauge group; however, it will appear in the broken case.

As mentioned, it is expected that for theories that can be obtained by softly breaking ${\cal N}=2$ to  ${\cal N}=1$ the genus zero contribution to the gravitational F-terms can be adsorbed by a field redefinition; on the other hand, in this paper we found that this is not necessarily true for theories that cannot be obtained in this way. It could be interesting then to precise in which kind of theories this redefinition can be done.

\noindent{\bf Acknowledgments} 

We would like to thank for useful discussions Justin David, Edi Gava and Kumar Narain. We thank also Luca Mazzucato.

\end{document}